# Isothermal anisotropic magnetoresistance in antiferromagnetic metallic IrMn

R. Galceran[1], I. Fina[1,2,*], J. Cisneros-Fernández[1], B. Bozzo[1], C. Frontera[1], L. López-Mir[1], H. Deniz[3], K.-W. Park[4], B.-G. Park[4], Ll. Balcells[1], X. Martí[5], T. Jungwirth[5,6], B. Martínez[1]

[1]Institut de Ciència de Materials de Barcelona (CSIC), Campus de Bellaterra, 08193 Bellaterra, Spain.

[2]Catalan Institute of Nanoscience and Nanotechnology (ICN2), CSIC and The Barcelona Institute of Science and Technology, Campus UAB, Bellaterra, 08193 Barcelona, Spain

[3]Max Planck Institute of Microstructure Physics, Weinberg 2, D-06120 Halle (Saale), Germany

[4]Department of Materials Science and Engineering, KAIST, Daejeon 305-701, Republic of Korea

[5]Institute of Physics ASCR, v.v.i., Cukrovarnicka 10, 162 53 Praha 6, Czech Republic

[6]School of Physics and Astronomy, University of Nottingham, Nottingham NG7 2RD, United Kingdom

Antiferromagnetic spintronics is an emerging field; antiferromagnets can improve the functionalities of ferromagnets with higher response times, and having the information shielded against external magnetic field. Moreover, a large list of aniferromagnetic semiconductors and metals with Néel temperatures above room temperature exists. In the present manuscript, we persevere in the quest for the limits of how large can anisotropic magnetoresistance be in antiferromagnetic materials with very large spin-orbit coupling. We selected IrMn as a prime example of first-class moment (Mn) and spin-orbit (Ir) combination. Isothermal magnetotransport measurements in an antiferromagnetic-metal(IrMn)/ferromagnetic-insulator thin film bilayer have been performed. The metal/insulator structure with magnetic coupling between both layers allows the measurement of the modulation of the transport properties exclusively in the antiferromagnetic layer. Anisotropic magnetoresistance as large as 0.15 % has been found, which is much larger than that for a bare IrMn layer. Interestingly, it has been observed that anisotropic magnetoresistance is strongly influenced by the field cooling conditions, signaling the dependence of the found response on the formation of domains at the magnetic ordering temperature.

---

*ignasifinamartinez@gmail.com

**Introduction**

During the recent years, antiferromagnetic (AF) materials[1,2] have been proposed as an alternative candidate for spintronic applications[3] in substitution of the traditional ferromagnets (FM). They would present the advantage of being robust against the presence of large magnetic fields at room temperature, because they do not show net magnetic moment to be modified. Moreover, compared to FM materials they present some interesting advantages: i) they do not generate stray fields thus, enabling further downscaling of the bit size in magnetic memories, ii) they are much more abundant in nature[4], and iii) faster switching than in FM has been predicted[5] and experimentally observed.[6] Ways to read/write on AF are currently under development.[7,8] Reading AF materials is not a real bottle-neck for their use because simple electrical-reading strategies similar to those used on FM materials can be used in AFs. The simplest electrical-reading technique is anisotropic magnetoresistance (AMR). AMR is an even function of the microscopic magnetic moment vector and depends on the relative orientation between the spin-axis alignment and the measuring current direction (*j*). Therefore, AMR is useful to infer perpendicular magnetic states in AFs. Indeed AMR has been used to obtain different resistive states corresponding to different magnetic states in AF semiconductor ($Sr_2IrO_4$)[9] and in AF metal (FeRh).[10] Similarly, tunneling anisotropic magnetoresistance can be used to infer magnetic states in AF.[11,12] Regarding the writing procedures, three methods have been demonstrated to be effective: i) exchange bias,[9,11,12] ii) heat assisted magnetic recording,[10] and most recently iii) electric spin transfer torque.[13] In the present work we will make use of the former, exchange bias,[14,15] by using a bilayer system where a FM layer adjacent to an AF material drags the magnetic moment alignment in the latter through the so-called exchange spring effect.

As mentioned, reports regarding AMR measurements on AF intrinsically modifying their microscopic magnetic moment vector are limited to $Sr_2IrO_4$,[9] CuMnAs[13] and non-isothermally in FeRh,[10] semiconductors and the metallic examples. However, to our best knowledge, isothermal AMR measurements on a metal have not been reported before. IrMn is a relevant AF material because it is often part of spin valve elements in commercial devices, and magnetic exchange coupling between AF IrMn and a FM has already been observed via Hall measurement,[16] permitting, in principle, to measure the intrinsic AMR in IrMn exploiting this magnetic interaction.

In the present work, we focus on the AMR effect in an IrMn layer grown on top of an insulating FM ($La_2CoMnO_6$) and magnetically coupled to it. Therefore, the unique metallic part in the bilayer is the AF IrMn layer. This distinct measurement configuration allows us to perform in-plane transport measurements and infer a sizeable AMR response (~0.15 %) in a metallic AF (IrMn) using exchange coupling. This value is one order of magnitude larger than that obtained measuring a bare IrMn film thus, confirming the intrinsic nature of the effect and the coupling between FM and AF layers.

**Results and discussion**

The characterized AF-metal (IrMn-2nm)/FM-insulator ($La_2CoMnO_6$, LCMO) bilayer is sketched in figure 1(a) together with a HAADF STEM image. The HAADF STEM image makes evident the good quality of the LCMO (further characterization can be found in Supplementary Figure S1) layer and the sharp contrast between the IrMn and LCMO layers with no signature of cationic intermixing. The polycrystalline nature of IrMn can be inferred from the Supplementary Figure S2. Figures 1(b,c) summarize the magnetic properties of the LCMO/IrMn bilayer. Figure 1(b) shows the temperature dependence of the magnetization recorded while cooling the sample under 100 mT field along [100]LCMO (in-plane) direction. From the figure it can be inferred that the magnetization increases while decreasing the temperature signaling the Curie temperature ($T_C$) of LCMO at 250 K. M(H) magnetization loop recorded at 5 K along the [100]LCMO (in-plane) is also shown in figure 1(c). From the figure, a coercive field of $H_C$ = 1 T is estimated. Note also that M(H) curve is still hysteretic up to about 3 T.

The temperature dependence of the resistance of the whole stack, measured by using the contact geometry shown in the sketch of figure 2(a), and that of two control samples [a bare IrMn film grown on a Si substrate (Si/IrMn) and a bare LCMO one] are shown in figure 2(b). The rapid increase of resistance of the LCMO layer (in black) while decreasing the temperature makes evident its insulating nature. In comparison, the resistance of the IrMn film (red line) is much less (extrapolation suggests around 9 orders of magnitude of difference at low temperatures), and comparable to that of LCMO/IrMn bilayer. Note that the resistance of IrMn slightly increases while decreasing the temperature; this effect can be ascribed to the low dimension of IrMn layer. The displayed transport characterization allows concluding that the overall resistance measured in the LCMO/IrMn bilayer can be ascribed to the IrMn layer.

Now we focus on magnetotransport characterization performed on the LCMO/IrMn stack and on an IrMn control film (Si/IrMn). Archetypical anisotropic magnetoresistance (AMR=$(R(\Phi)-R(\Phi=0))/R(\Phi=0)$) measurements have been performed by rotating the applied magnetic field by an angle $\Phi$ in the plane of the sample as sketched in figure 2(c), where $\Phi$ is defined as the angle contained in the plane of the film with $\Phi = 0°$ at applied field perpendicular to the measurement current. Regarding LCMO, the magnetic field is applied along the [100] direction at 0° and along the [010] at 90°. The applied magnetic field (3 T) is well above the coercive field, thus ensuring that LCMO magnetization is aligned parallel the applied magnetic field. The central result of our work is shown in figure 2(d): an AMR response as large as 0.15 % is measured on the IrMn/LCMO bilayer at 5 K after a field cooling process (FC) from high temperature (400 K) in a magnetic field of 2T applied along the [100] direction. In contrast, the bare IrMn film does not show any relevant dependence on the applied field direction. The contrast between both measurements indicates that in the bilayer system the LCMO magnetic moment rotates following the applied field, and this rotation promotes the rotation of the magnetic spin alignment of the IrMn layer, allowing us to detect the AMR response of IrMn. Note also that the AMR is asymmetric and more pronounced for one direction of the magnetic field (with $\Theta$ near to 90°). However, opposite asymmetry in the AMR is observed when the field cooling process is performed in the same conditions but with H pointing along the [010] direction. AMR response after a zero-field-cooling process (ZFC, empty squares) is significantly smaller and does not exhibit the archetypical $\cos^2\Theta$ shape. The relevance of these results is twofold. Not only they confirm the existence of a magnetic coupling between the IrMn and the LCMO layers but also the distinct shape of the measured AMR clearly evidences the relevance of the field cooling conditions on the magnetic coupling. Complementary experiments on equivalent sample with a 10 nm thick IrMn layer instead of 2 nm show that the AMR vanishes (Supplementary Figure S3), revealing that the effect is confined at the interface as expected for an exchange spring effect. The found variation of AMR (0.15 %) is smaller than that found in antiferromagnetic FeRh films,[10] where the used methodology was not isothermal. However, in the present case the value can be largely influenced by the fact that the IrMn magnetic order is only partially dragged by LCMO. The weak coupling is also revealed by the temperature dependence of the AMR (shown below) which shows rapid vanishing while increasing the temperature.

The existence of magnetic coupling between both layers is also signaled by the shift in the magnetic hysteresis loops recorded at 5 K, performed after cooling the LCMO/IrMn bilayer with a magnetic field applied along opposite directions. These loops are shown in figure 3(a), the large coercive field of LCMO makes difficult to distinguish the presence of exchange bias. Zoomed image shown in figure 3(b), allows identifying a sizeable difference when comparing both loops. For the sake of clarity, M(H) curves have been fitted to the expression $M=M_0 \cdot \tanh[(H-H_C)/\delta H] + \chi_{PM} \cdot H$,[17] where $M_0$ accounts for the switchable magnetization, $H_C$ for the magnetic coercive field, $\delta H$ for the width of the magnetic susceptibility, and $\chi_{PM}$ for the paramagnetic contribution. Lines through data point correspond to the fitted data in figures 3(a,b). The exchange bias field [$(H_{EB} = H_C^+ - H_C^-)/2$] after the fitting is plotted in figure 3(c) (for the complete list of the values extracted from the fits see Supplementary Table S1). The $H_{EB}$ magnitude ($H_{EB} \approx 1.5$ mT) lead to an interface energy ($\Delta E = H_E t_{FM} M_{FM}$) of $\approx 0.01$ erg/cm$^2$, which is considered low, but in agreement with previous estimations for IrMn films grown on other ferromagnetic materials.[14]

The dependence of the AMR response on the intensity of the applied magnetic field is shown in figure 4(a). The rotation of the sample with respect to the field direction at H=0 T provides the noise level of the experiment. Increasing the field to H=1 T, sizeable AMR values, yet scattered, are measured. AMR gradually increases while increasing the magnetic field, as evidenced also by the AMR map of figure 4(b). In the latter, it is interesting to note that the AMR saturates near 3 T and onwards. The saturation can be correlated with the fact that even though $H_c$ is about 1 T as shown by M(H) loops of figure 1(c), the irreversibility field extends up to 3 T. On the other hand, it is also worth mentioning that AMR strongly depends on the orientation of the applied magnetic field with respect to the sample plane (see fig. 4(c)). AMR response is stronger in the in-plane configuration (IP) [equivalent to figure 2(d)] than in the out-of-plane one (OOP), where the magnetic field is rotated from a direction perpendicular to the current ($\alpha$=0) to perpendicular to the plane of the film [$\alpha$=90º, where $\alpha$ is the angle containing the directions perpendicular to the current in the plane of the film and the normal to the film as sketched in figure 2(c)]. The non-zero value of both the IP- and the OOP- AMR indicates that the magnetization of LCMO drags somehow the antiferromagnetic domains of IrMn in both directions; however, if the coupling is smaller when the magnetization is pointing along OOP-axis.

Finally, the temperature dependence of the AMR (at 3 T) is shown in figure 4(d). The dependence of AMR on temperature shows that AMR vanishes at 15 K well below $T_C$ and $T_N$.[11] The rapid vanishing of the AMR with temperature signals that the magnetic coupling between both layers is rather weak. This result, complemented by the observed asymmetric behavior of the AMR response depending on the field cooling conditions (figure 2d), allows safely discarding uncompensated spins, and spin hall effect[18,19] as the main sources of the observed effect, because these contribution are expected to remind as far as the insulating layer is ferromagnetically ordered (up to 250 K).

**Summary and Conclusions**

We report sizeable AMR response of 0.15 % in LCMO/IrMn bilayer system that is uniquely ascribed to IrMn, since the resistivity of LCMO layer is about 9 orders of magnitude above that of IrMn layer.

This is observed in contrast with the negligible effect measured in a stand-alone IrMn layer, being former result the first isothermal measurement of AMR in a metallic antiferromagnetic material. The effect is observed owing to the existence of a magnetic exchange coupling between the LCMO and IrMn adjacent layers via exchange spring effect, also revealed by direct magnetometric data. The dependence of AMR response on the field cooling conditions and temperature together with the asymmetric behavior of the AMR response depending on the field cooling conditions, allows safely discarding uncompensated spins as the source of the observed behavior.

**Methods**

*Sample growth*

An IrMn layer (2 nm) was grown on top of a ~15.8 nm thick $La_2CoMnO_6$ (LCMO) layer, both prepared in-situ by RF magnetron sputtering. The LCMO layer was c-oriented, epitaxial and fully strained (see ref. 20 and Supplementary Figure S1 for further structural characterization) and it was grown on top of a (001)-oriented $SrTiO_3$ substrate. The stacking sequence is shown in Fig. 1(a). Pervious works on IrMn films of similar thickness make evident using heat capacitance experiments the AF nature of the IrMn-layer[11] and that magnetic coupling between IrMn and adjacent FM layer is present via an exchange spring effect even at the last IrMn atomic layers.[15]

*TEM characterization*

TEM images have been performed by using FEI TITAN 80–300 operated at 300 kV.

*Magnetotransport characterization*

The resistance of the bilayer has been measured using a 4-points contact method. The contacts are placed on top of the structure as shown by figure 2(a). The excitation current was set to 100 μA. The magnetic characterization was performed by using a SQUID magnetometer (Quantum Design). Transport properties were measured in a Physical Properties Measurement System (PPMS, Quantum Design) using a rotating sample holder allowing a precise control of magnetic field (magnitude and direction) and the temperature. Temperature and magnetic field were varied at a constant rate of 2 K/min and 180 Oe/s, respectively. Raw measurements of the magnetization vs applied magnetic field were corrected by subtracting the diamagnetic contribution from the substrate.

**References**

1  MacDonald, A. & Tsoi, M. Antiferromagnetic metal spintronics. *Philosophical Transactions of the Royal Society A: Mathematical, Physical and Engineering Sciences* **369**, 3098-3114 (2011).
2  Gomonay, E. & Loktev, V. Spintronics of antiferromagnetic systems (Review Article). *Low Temperature Physics* **40**, 17-35 (2014).
3  Shick, A. B., Khmelevskyi, S., Mryasov, O. N., Wunderlich, J. & Jungwirth, T. Spin-orbit coupling induced anisotropy effects in bimetallic antiferromagnets: A route towards antiferromagnetic spintronics. *Physical Review B* **81**, 212409 (2010).
4  Goodenough, J. B. Magnetism and chemical bond. (Interscience publishers, 1963).
5  Železný, J. *et al.* Relativistic N\'eel-Order Fields Induced by Electrical Current in Antiferromagnets. *Physical Review Letters* **113**, 157201 (2014).
6  Kimel, A., Kirilyuk, A., Tsvetkov, A., Pisarev, R. & Rasing, T. Laser-induced ultrafast spin reorientation in the antiferromagnet TmFeO3. *Nature* **429**, 850-853 (2004).
7  Jungwirth, T., Marti, X., Wadley, P. & Wunderlich, J. Antiferromagnetic spintronics. *Nat Nano* **11**, 231-241, doi:10.1038/nnano.2016.18 (2016).
8  Marti, X., Fina, I. & Jungwirth, T. Prospect for antiferromagnetic spintronics. *Magnetics, IEEE Transactions on* **51**, 1-4 (2015).
9  Fina, I. *et al.* Anisotropic magnetoresistance in an antiferromagnetic semiconductor. *Nat Commun* **5**, 4671, doi:10.1038/ncomms5671 (2014).
10  Marti, X. *et al.* Room-temperature antiferromagnetic memory resistor. *Nature Materials* **13**, 367 (2014).
11  Petti, D. *et al.* Storing magnetic information in IrMn/MgO/Ta tunnel junctions via field-cooling. *Applied Physics Letters* **102**, 192404 (2013).
12  Park, B. *et al.* A spin-valve-like magnetoresistance of an antiferromagnet-based tunnel junction. *Nature Materials* **10**, 347-351 (2011).
13  Wadley, P. *et al.* Electrical switching of an antiferromagnet. *Science* **351**, 587-590, doi:10.1126/science.aab1031 (2016).
14  Nogués, J. & Schuller, I. K. Exchange bias. *Journal of Magnetism and Magnetic Materials* **192**, 203-232 (1999).
15  Martí, X. *et al.* Electrical measurement of antiferromagnetic moments in exchange-coupled IrMn/NiFe stacks. *Physical Review Letters* **108**, 017201 (2012).
16  Zhang, X. & Zou, L. Planar Hall effect in Y3Fe5O12/IrMn films. *Applied Physics Letters* **105**, 262401 (2014).
17  Takacs, J. A phenomenological mathematical model of hysteresis. *COMPEL-The international journal for computation and mathematics in electrical and electronic engineering* **20**, 1002-1015 (2001).
18  Isasa, M. *et al.* Spin Hall magnetoresistance at Pt/CoFe2O4 interfaces and texture effects. *Applied Physics Letters* **105**, 142402 (2014).
19  Mendes, J. *et al.* Large inverse spin Hall effect in the antiferromagnetic metal Ir 20 Mn 80. *Physical Review B* **89**, 140406 (2014).
20  Galceran, R. *et al.* Engineering the microstructure and magnetism of La2CoMnO6− δ thin films by tailoring oxygen stoichiometry. *Applied Physics Letters* **105**, 242401 (2014).

## Acknowledgments

We acknowledge financial support from the Spanish MINECO (MAT-2015-71664-R, MAT2015-73839-JIN) and FEDER program. We also acknowledge support from the ERC Advanced grant no. 268066, from the Ministry of Education of the Czech Republic Grant No. LM2011026, from the Grant Agency of the Czech Republic Grant no. 14-37427. ICMAB-CSIC authors acknowledge financial support from the Spanish Ministry of Economy and Competitiveness, through the "Severo Ochoa" Programme for Centres of Excellence in R&D (SEV- 2015-0496). R.G. thanks the Spanish MINECO for the financial support through the FPI program. I.F. acknowledges the Beatriu de Pinós postdoctoral scholarship (2011 BP-A_2 00014) from AGAUR-Generalitat de Catalunya. B.-G.P. acknowledges support from NRF of Korea funded by the Ministry of Science, ICT & Future Planning (NRF-2014R1A2A1A11051344). We finally thank Helmholtz-Zentrum Berlin for the allocation of synchrotron radiation beamtime, and Dr. Daniel M. Többens for his kind assistance during data collection.

## Author contributions

I.F., X.M., T.J., and B.M. jointly designed and conceived the experiments. The samples were grown and characterized by R.G., J.C-F, C.F., L.L.-M., K.-W.P., B.-G.P, and L.B.. R.G., H.D. performed TEM characterization. R.G., I.F., and B.B. conducted the magnetotransport experiments. I.F. wrote the paper. All authors discussed data and commented on the paper.

## Additional Information

*Competing financial interests*

The authors declare no competing financial interests.

## Supplementary Material

Supplementary Information: Available at…

## Captions

**Figure 1. Materials and magnetic properties.** (a) Sketch and TEM cross-section of antiferromagnetic-metallic (IrMn, 2 nm)/ferromagnetic-insulator (La$_2$CoMnO$_6$, LCMO) bilayer (b) Magnetization vs. temperature measured with H = 100 mT. (c) Magnetization vs. applied field measured at 5 K with H. In (a) and (b) H is applied in- the plane of the sample (//[100]STO).

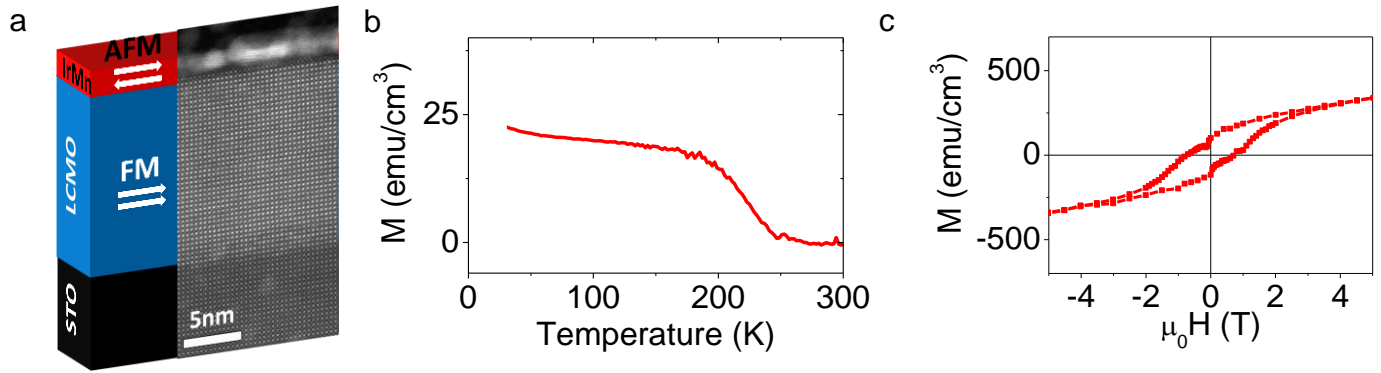

**Figure 2. Transport and magnetotransport characterization.** (a) Sketch of the electrical contact configuration for the STO/LCMO(20 nm)/IrMn(2 nm). Equivalent to the one used in the Si/IrMn sample. (b) Temperature dependence of the absolute measured resistance for LCMO films with and without IrMn on top . Resistivity for IrMn at 5 K, temperature for the sample grown on SiOx and the one on LCMO is near 700 μOhms·cm. (b) Sketch of the applied magnetic field orientation with respect to the measurement current (d) AMR = (R(Φ)-R(Φ=0))/R(Φ=0) measurements at 5 K and μ₀H = 3 T for STO/LCMO/IrMn and Si/IrMn samples recorded after field-cooling the samples at 2 T Φ=0º and Φ=90º indicate the field cooling direction perpendicular and parallel to the current, respectively.

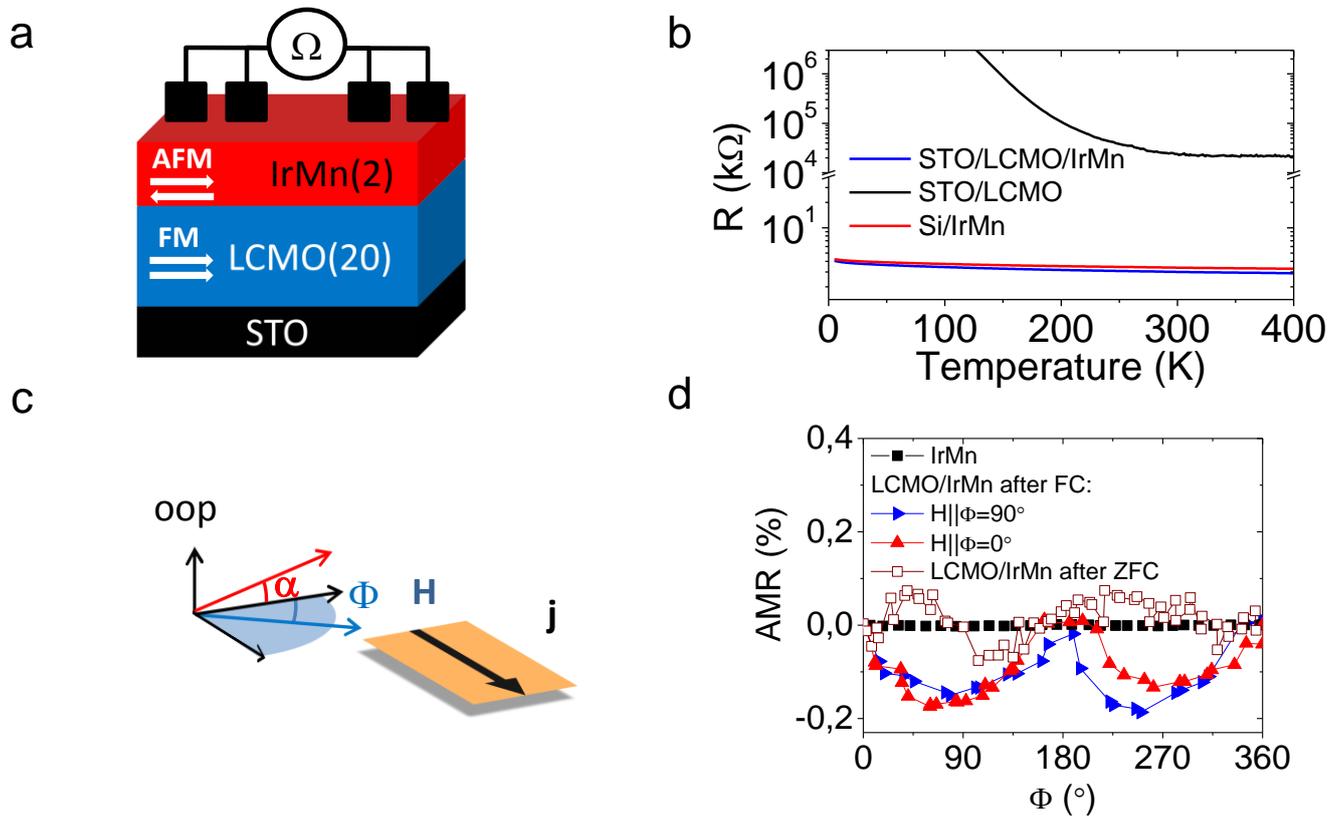

**Figure 3. Exchange magnetic coupling effect.** (a) Magnetization versus applied magnetic field loops recorded after field cooling the samples along opposite directions. (b) Zoom of the (a) panel. In a and b, lines though data points correspond to the fitting of the expression $M=M_0\cdot\tanh[(H-H_C)/\delta H] + \chi_{PM}\cdot H$. (c) $H_{EB}$ extracted from 4 different hysteresis loops recorded upon successive field cools along opposite directions is plotted.

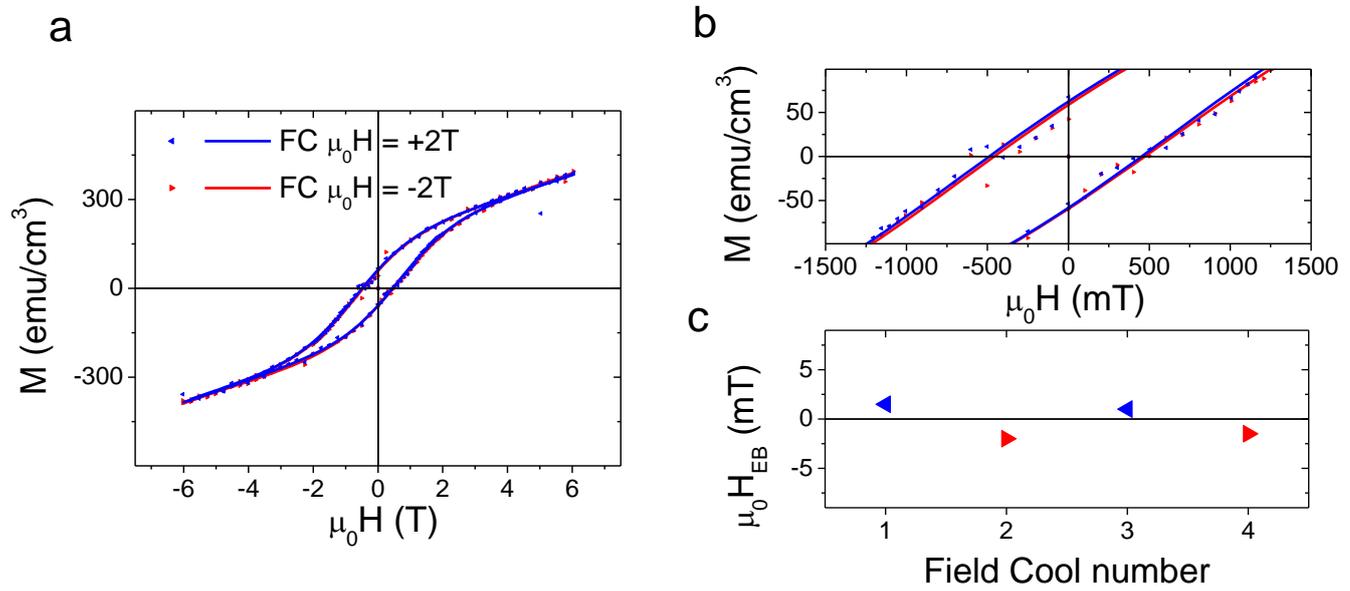

**Figure 4. Temperature and magnetic field dependence of anisotropic magnetoresistance.** (a) AMR measurements at 5 K and various fields for STO/LCMO/IrMn sample. (b) Map of AMR measurements at 5 K and various fields IrMn. (c) Dependence of the AMR on the plane of rotation of the magnetic field at 5 K. IP is the plane of the sample [with $\Phi$ angle as defined in figure 2(b)] and the OOP is the plane containing the directions of the current and the normal to the film [with $\alpha$ angle as defined in figure 2(b)]. (d) AMR recorded at different temperatures at 3T.

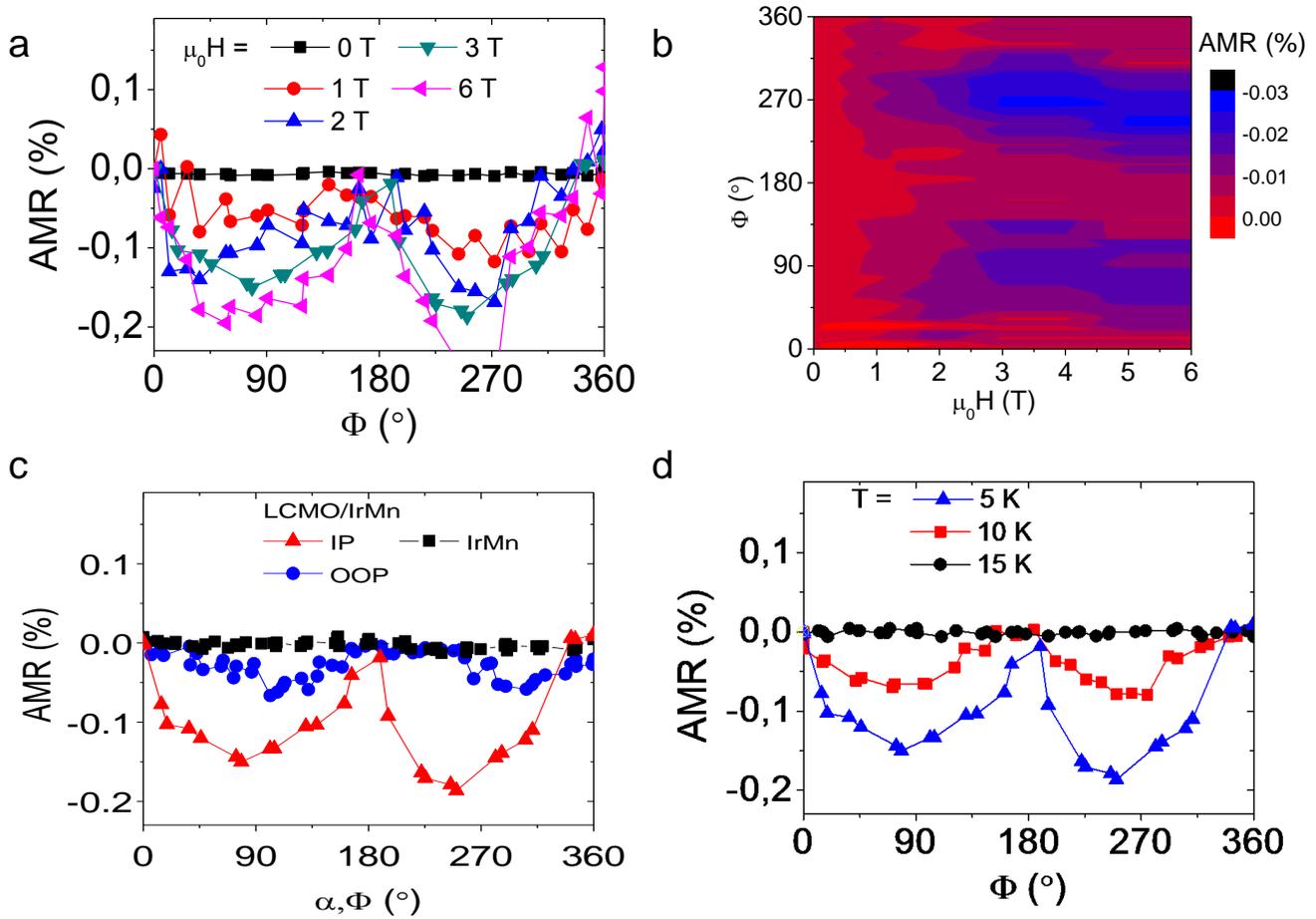

# Supplementary Information on: Isothermal anisotropic magnetoresistance in antiferromagnetic metallic IrMn

R. Galceran[1], I. Fina[1,2,†], J. Cisneros-Fernández[1], B. Bozzo[1], C. Frontera[1], L. López-Mir[1], H. Deniz[3], K.-W. Park[4], B.-G. Park[4], Ll. Balcells[1], X. Martí[5], T. Jungwirth[5,6], B. Martínez[1]

## Figure S1. Detailed structural characterization of LCMO film

Figure S1(a) shows the reciprocal space map of (103) diffraction peak of a STO/LCMO film. This shows the high epitaxial character of the growth and that film is fully strained. The good quality of the film can be also observed in the zoom of the TEM cross-section image of the LCMO film shown by figure S1(b).

Synchrotron X-ray diffraction of figure S1a has been done at KMC2 beamline (BESSY-II at Helmholtz-Zentrum, Berlin), which is equipped with a two dimensional General Area Detector Diffraction System.

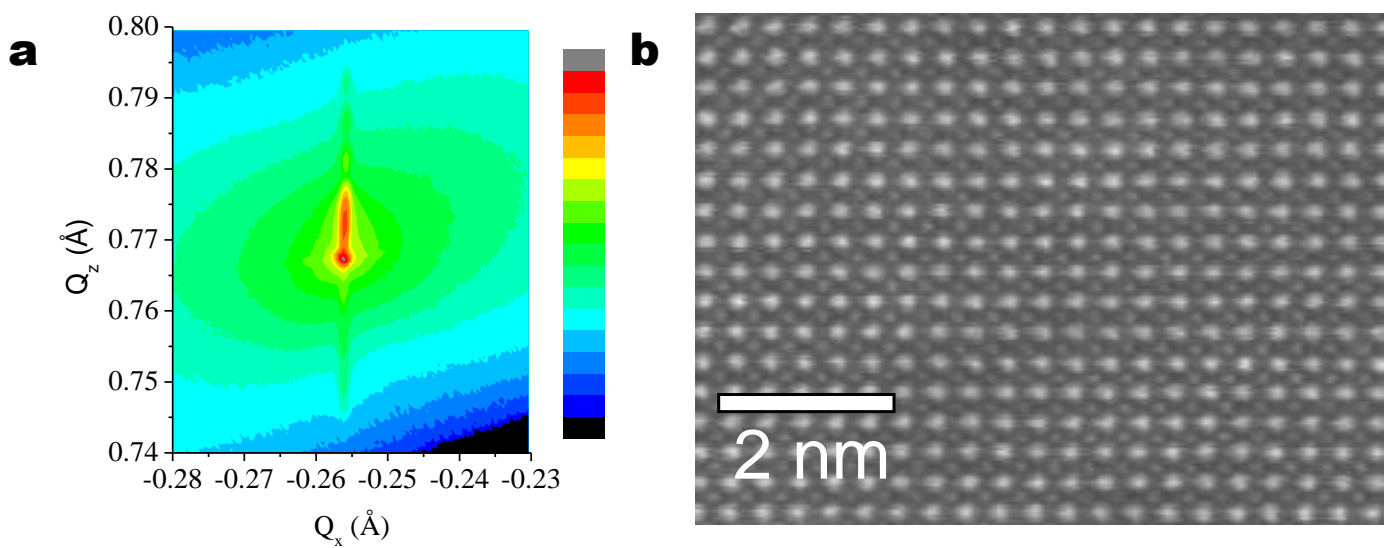

## Figure S2. Transition electron microscopy image of the IrMn layer

The zoom of the TEM cross-section of antiferromagnetic-metallic (IrMn, 2.2 nm)/ferromagnetic-insulator (La2CoMnO6, LCMO) bilayer, shown in figure S2, reveals that the IrMn layer has a grainy morphology. Several grains are crystalline (signalled by arrows). Other grains show no- clear crystallinity, because of the non-proper orientation of the crystalline planes with respect to electron beam, signalling the polycrystalline nature of IrMn layer.

---

†ignasifinamartinez@gmail.com

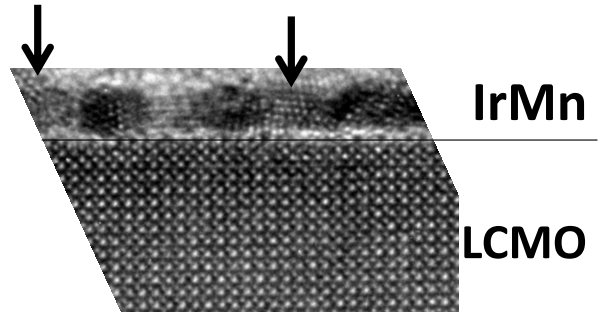

## Figure S3. Anisotropic magnetoresistance of IrMn(10 nm)/LCMO bilayer

In figure S3, we show the AMR = (R($\Phi$)-R($\Phi$=0))/R($\Phi$=0) measurements performed at 5 K under a magnetic field of 3 T for a IrMn(10 nm)/LCMO film (after field-cooling the samples at 2 T $\Phi$=0º).. For comparison, we have also included the result for the IrMn(2.2 nm)/LCMO sample of the main manuscript (after field-cooling the samples at 2 T $\Phi$=0º and $\Phi$=90º). It can be observed that no-sizeable signal is observed for the 10 nm IrMn film.

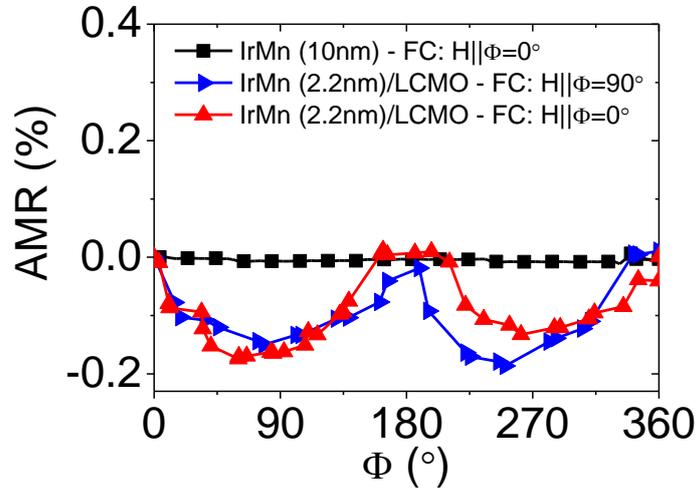

## Table S1. Fitting results for the exchange bias

Table S1 shows the values of the fits obtained by the expression $M_0 \cdot \tanh[(H-H_C)/\delta H] + \chi_{PM}$ after field cooling the sample with magnetic field applied along opposite directions as indicated in the first raw.

| | Field cooling: | Negative 1 | Positive 1 | Negative 2 | Positive 2 |
|---|---|---|---|---|---|
| | | Value | Value | Value | Value |
| Increasing H branch | $M_0$ (emu·10$^{-5}$) | 7.2 | 6.7 | 7.2 | 6.3 |
| | $H_c$ (mT) | 623 | 618 | 640 | 627 |
| | $\delta H$ (mT) | 1890 | 1739 | 1886 | 1645 |
| | $\chi$ (emu·10$^{-13}$/T) | 3.56 | 3.44 | 3.54 | 3.40 |
| Decreasing H branch | $M_0$ (emu·10$^{-5}$) | 6.3 | 7.1 | 6.4 | 6.4 |
| | $H_c$ (mT) | -620 | -622 | -638 | -630 |
| | $\delta H$ (mT) | 1717 | 1883 | 1677 | 1677 |
| | $\chi$ (emu·10$^{-13}$/T) | 3.36 | 3.49 | 3.38 | 3.38 |
| | $H_{EB}$ (mT) | 1.5 | -2 | 1 | -1.5 |